# Chapter 8

# Probing light by matter: Implications of complex illumination on ultrafast nanostructuring

**Camilo Florian[1, a], Xiaohan Du[1, 2, b], Prof. Craig B. Arnold[1, 2, c]**

[1] Princeton Institute for the Research and Technology of Materials (PRISM), Princeton University, 70 Prospect Av, 08540, Princeton, NJ, USA.

[2] Mechanical and Aerospace Engineering (MAE), Princeton University, 50 Olden Street, 08540, Princeton, NJ, USA

[a] camilo.florian@princeton.edu, [b] xiaohand@princeton.edu, [c] cbarnold@princeton.edu

**Abstract:**

Pushing the limits of precision and reproducibility in ultrafast laser-based nanostructuring requires detailed control over the properties of the illumination. Most traditional methods of laser-based manufacturing rely on the simplicity of Gaussian beams for their well-understood propagation behavior and ease of generation. However, a variety of benefits can be obtained by moving beyond Gaussian beams to single or multiple tailored beams working towards optimal spatial and temporal control over the beam profiles. In this chapter, we center our attention on methods to generate and manipulate complex light beams and the resulting material interactions that occur in response to irradiations with these non-traditional sources. We begin with a discussion on the main differences between Gaussian and more complex light profiles, describing the mechanisms of phase and spatial control before narrowing the discussion to approaches for spatial structuring associated with materials processing with ultrashort laser pulses. Such structuring can occur in both far-field propagating architectures, considering rapidly varying spatial profiles generated mechanically or optically, as well as near-field, non-propagating beams associated with plasmonic and dielectric systems. The chapter emphasizes some of the unique abilities of complex light to shape materials at the nanoscale from a fundamental perspective while referencing potential applications of such methods.

processing, dual laser additive manufacturing, direct laser-interference patterning, laser-induced forward transfer, resonant focal scanning, TAG lens.

# Table of contents

# 1 Introduction: Traditional illumination versus spatially structured light

*The Gaussian Beam*

The most commonly used laser beam shape for many micromachining applications is the Gaussian distribution. It is not a surprise since the vast majority of laser cavities are designed to operate at or near the fundamental transverse electromagnetic mode (TEM00) which corresponds to a Gaussian. One clear advantage from the mathematical standpoint is that its Fourier transform is also Gaussian, meaning that it is possible to keep a Gaussian distribution even after the beam passes through focusing elements such as lenses and objectives. Therefore, its implementation in laser processing systems can be done in most cases without the need for beam shaping techniques [Duocastella, 2012]. With regard to materials processing applications, there are three general parameters that are considered important due to their influence on the final features imprinted by the laser radiation. The first one, known as the **confocal parameter,** is the distance range along the propagation axis where the energy can be confined. The second, the **beam divergence,** depends on how the beam changes along the propagation axis. Finally, it is important to know the **minimum spot size** where the energy can be localized. For obvious reasons, when micro- or nano-machining is desired, these beam dimensions play an important role in the final modification that is ultimately achievable in the irradiated material. These parameters can be deduced from the mathematical expression that describes an electric field propagating along the Z axis (*E(r,z)*) of a linearly polarized Gaussian beam distribution in the Z direction as shown in Eq. 1-1 [Silfvast, 2004]:

$$E(r,z) = E_0 \frac{w_0}{w(z)} \exp\left(\frac{-r^2}{w(z)^2}\right) \exp\left[-i\left(kz + k\frac{r^2}{2R(z)} - \varphi(z)\right)\right], \quad \text{Eq. 1-1}$$

where $z$ is the position on the propagation Z axis, $w_0$ is the waist radius, $i$ is the unit imaginary number, $r$ is the radial distance from the center (Eq. 1-2), and $k$ is the wave number (Eq. 1-3).

$$r = \sqrt{x^2 + y^2}, \quad \text{Eq. 1-2}$$

$$k = \frac{2\pi}{\lambda}. \quad \text{Eq. 1-3}$$

The beam radius along the axis propagation, *w(z)* can be expressed as a function of a parameter known as Rayleigh length:

$$w(z) = w_0\sqrt{1 + \left(\frac{z}{z_r}\right)^2}. \quad \text{Eq. 1-4}$$

The Rayleigh length is the distance from the position of the minimum beam waist, $w_0$, along the propagation axis until which the cross-section of the beam doubles, or the beam radius increases by a factor $\sqrt{2}$:

$$z_R = \frac{\pi w_0^2}{\lambda}. \quad \text{Eq. 1-5}$$

The **confocal parameter** is defined as two times the Rayleigh length, $2z_R$, and is usually used to indicate the spatial range along the propagation axis where the energy can be confined.

The radius of curvature of the wave front for distances larger than the Rayleigh length (in the far-field region) is defined as:

$$R(z) = z\left[1 + \left(\frac{z_R}{z}\right)^2\right]. \quad \text{Eq. 1-6}$$

Finally, an additional phase shift that occurs when a Gaussian beam is being focused is known as the Gouy phase, and it can be described as:

$$\varphi(z) = \tan^{-1}\left(\frac{z}{z_R}\right). \quad \text{Eq. 1-7}$$

When the beam propagates from the focal plane further than the Rayleigh length (Eq. 1-5), i.e., in the far-field, a non-ideal Gaussian beam expands laterally far from the beam waist. The resulting angle difference measured from the position of the minimum beam waist, $w_0$, is known as the **beam divergence angle** $\theta$. In the case of Gaussian beams, the beam quality factor $M^2$ allows us to describe how far from an ideal Gaussian distribution a real laser beam is when it propagates to infinity. For ideal diffraction-limited Gaussian beams, $M^2 = 1$, while for real beams, $M^2 > 1$.

$$M^2 = \frac{\pi w_0 \theta}{\lambda}, \quad \text{Eq. 1-8}$$

$$\theta = \frac{M^2 \lambda}{\pi w_0}. \quad \text{Eq. 1-9}$$

Gaussian beams are characterized by having low divergence, which allows focusing the energy from these beams in very small volumes. A good approximation for the **minimum spot size**, $d_f$, that it is possible to achieve with a Gaussian beam that is being focused through a focusing lens with numerical aperture NA, can be expressed as:

$$d_f \approx 2\frac{\lambda M^2}{\pi NA}. \quad \text{Eq. 1-10}$$

Traditionally, the corresponding intensity distribution $I(r, z)$ for Gaussian beams is given by Eq. 1-11 of the modulus squared of the electromagnetic field (Eq. 1-1) with the impedance of the medium in which the field is propagating $\eta$ (for propagation in vacuum $\eta = \eta_0 \approx 377\ \Omega$). A plot of the intensity profile corresponding to a Gaussian beam is included in Table 1 at the end of the current Section 2.

$$I(r,z) = \frac{|E(r,z)|^2}{2\eta}.  \qquad \text{Eq. 1-11}$$

*Implications for materials processing*

The modification of materials with high spatial resolution typically requires high effective numerical aperture systems to confine a Gaussian beam into the tiniest possible voxel. For example, light from an infrared laser of wavelength 1027 nm with Gaussian distribution and beam quality factor $M^2 = 1.1$, focused through a lens system with NA = 0.55 (commonly found in 50× microscope objectives) is axially confined to around 6 µm (Eq. 1-5) and laterally to 1.43 µm (Eq. 1-10). When the irradiation is done at the surface level of a static sample, the resulting material modification will follow a crater-like morphology. For modifications in bulk, such as in the particular case of transparent materials, it is actually possible to produce a modification whose geometry will closely follow the voxel dimensions[1]. Figure 1 shows the different ablated morphologies created by a single shot of femtosecond laser for surface and bulk modification over a sample of poly-methyl methacrylate (PMMA). When the voxel of the focused beam is located close to the surface, only a superficial crater-like modification is produced. However, when the voxel is located slightly beneath the surface, the irradiation produces a modification in the bulk of the material following the dimensions and shape of the focused beam voxel.

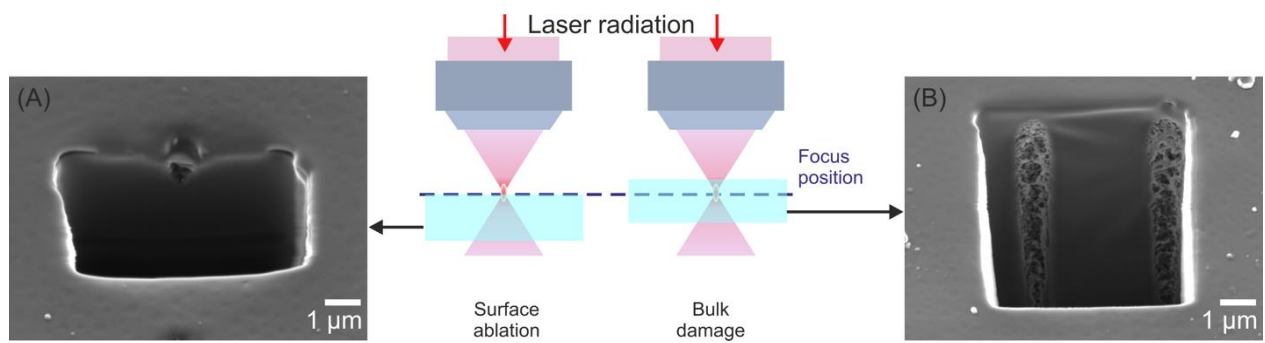

Figure 1. SEM images of laser irradiations from a single femtosecond laser pulse (1027 nm, 450 fs) focused through a NA = 0.55 microscope objective (A) at the surface and (B) slightly beneath the surface of a sample of poly-methyl methacrylate (PMMA). A focused ion beam was used to selectively ablate material to show the axial extend of the beam modification. The images were acquired with an incident angle of 45° to visualize the walls of the FIB ablated area [Florian, 2015]. Copyright (2015), with permission from Elsevier.

---

[1] The resulting modification size and shape will be strongly influenced by the light-matter interaction.

From an application point of view, it will be possible to modify a material with a focused Gaussian beam as far as the material surface is located within the focused voxel location, and the intensity there is adequate to modify it. Nevertheless, this is not always the case, especially for non-flat surfaces with high roughness, which are abundant in many materials processing industrial applications. The challenge here is maintaining an accurate alignment between the material of interest and the location of the focused voxel. Although this could be solved by using high accuracy positioning stages, they could considerably increase the overall production costs and extend the processing times. For this reason, modifications on the laser beam intensity and propagation constitute robust alternatives to extend the advantages of traditional Gaussian sources.

### *Structured light: beyond Gaussian beams*

Among the different variations that could be performed over a beam of light to increase the efficiency towards the micro- and nanofabrication of materials, we could classify the most widespread methods into two groups: (i) techniques that rely on spatial intensity distribution changes only (in far and near-field configurations) and (ii) time-dependent structured profiles that could modify the beam intensity distribution, its propagation and/or its interaction with matter.

In the first group, the objective is to **spatially modify the shape of the intensity profile** to meet the conditions for the final processing requirements. A good example widely used in materials processing applications is the *Top-hat beam* [Rung, 2013]. Its formation comes from a simple modification to a Gaussian beam where the final spatial distribution forms a homogeneous intensity plateau at the center. This plateau can modify materials, producing well-defined modification sizes, shapes and depths, while at the same time being practically independent of the laser energy used. It is particularly convenient when different material thicknesses need to be processed while maintaining the same modification dimensions. We will discuss different spatial beam shapes in the far-field in *Section 2* and near-field configurations in *Section 3* that offer advantages over Gaussian beams in terms of higher spatial resolution features or longer fabrication axial ranges.

In the second group, the **time-dependent structured profiles** consist of a rapid modification to the beam intensity profile, its propagation, or its interaction dynamics with matter. As an example of these profiles, the use of a tunable acoustic gradient of index (TAG) lens in a regular direct-laser writing system can induce rapidly varying intensity profiles over extended axial ranges, providing axially longer working distances compared with traditionally focused Gaussian beams at operational oscillation frequencies over ~140 KHz [Du, 2021]. The rapidly changing properties of the beam in time enables the use of these types of complex beams in advanced materials processing applications, as we will develop further in *Section 4*. Lastly, the chapter will close by providing an overview of the presented beam shaping techniques and their standpoints for the machining of materials towards the pursuit of extremely small scales.

## 2 Far field approaches

### 2.1 Top hat beams

One method for the generation of a Top-hat intensity distribution is the use of a diffractive mask placed in the propagation path of a Gaussian beam to form an Airy pattern that will be subsequently focused through a lens [Rung, 2013]. The distribution at the focal plane corresponds to a circular beam with a homogeneous intensity area forming a plateau. Contrary to a Gaussian beam distribution, the edges of this intensity distribution are sharp, forming the characteristic hat shape as it is illustrated in Figure 2. In comparison with Gaussian beams, the quality factor $M^2$ is usually over 10, which does not enable tight focusing or axial confinement (Eq. 1-5 and Eq. 1-10). Despite that, the advantages that can be retrieved from the homogeneous intensity plateau in materials processing applications could be exploited for laser cutting [Rung, 2013], laser annealing [Miyasaka, 1999], and pulsed laser deposition [Chrisey, 1994], to name a few. Also, this beam distribution is useful for several imaging applications, since the homogeneous intensity area can be exploited to illuminate a specific area of interest homogeneously [Rowlands, 2018]. Regarding the final modification of materials with this type of beam, Figure 2 includes a comparison that shows the typical formation of a crater-like modification from a Gaussian beam and the squared well-like modification from a Top hat distribution.

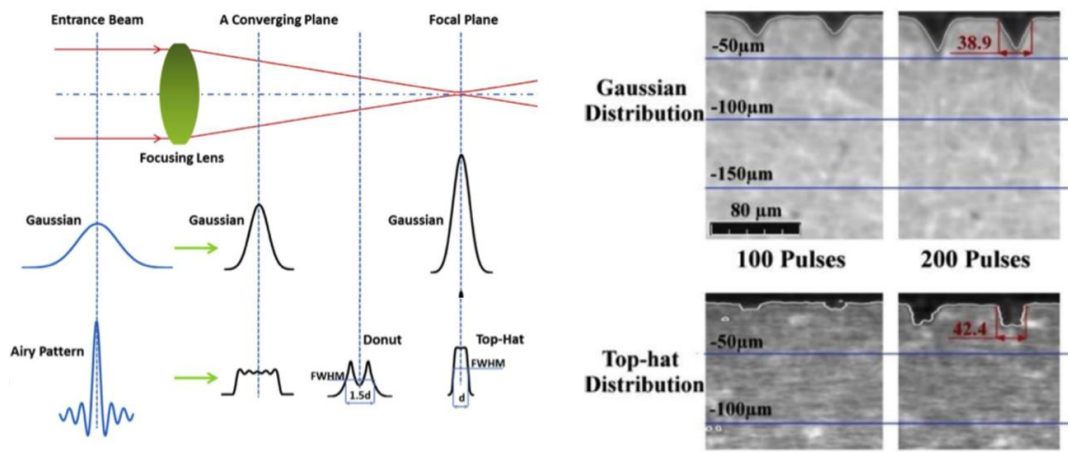

Figure 2. Gaussian and Airy patterns focused through a regular lens, producing Gaussian and Top-hat intensity distributions respectively at the focal plane. On the right, a side view optical image

shows the corresponding material ablation on a sample of silicon nitride, after the irradiation with 100 and 200 pulses. It can be seen how in the case of the Top-hat profile, the ablation follows the squared beam shape in comparison with the Gaussian counterpart. Reprinted from [Nasrollahi, 2020] Copyright (2020), with permission from Elsevier.

## 2.2 Annular and Vortex beams

Gaussian beams could also be modified to form *Annular beam distributions* that correspond to doughnut-shaped intensity profiles in which the intensity peak is arranged to form a ring with a minimum at the center. Similarly, as for the production of a Top-hat beam, when an Airy pattern is focused through a focusing lens, at positions before the focal plane, an Annular intensity distribution can be formed, as illustrated in Figure 2 [Nasrollahi, 2020]. Commonly, applications for Annular beams can be found in illumination systems, since it is possible to induce changes over a ring-shaped region while keeping the center unaffected. This effect is exploited in the super-resolution microscopy technique STED [Hell, 1994]. For materials processing, the machining application should require the fabrication of ring-like modifications like the one shown in Figure 3. An example can be found in the propelling of materials induced by blister-assisted laser-induced forward transfer (BA-LIFT) [Suhara, 2020]. In that particular case, the generation of a ring-like modification at the interface between a polymer layer and a transparent substrate, generates a protrusion that propels the liquid forward over longer than usual distances without compromising the spatial resolution of the transferred material.

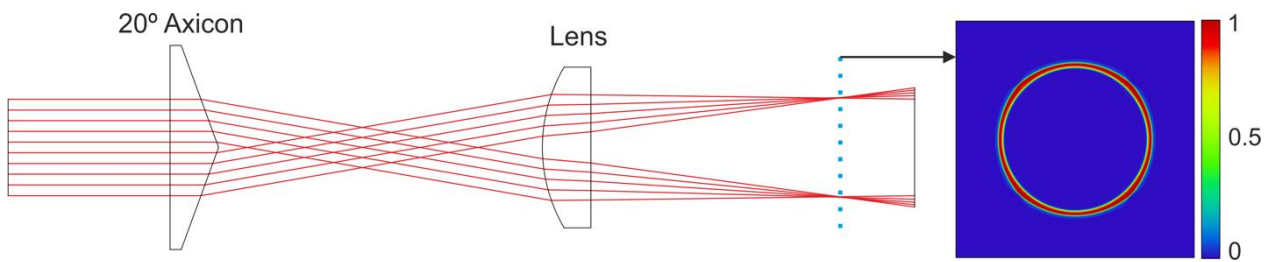

Figure 3. Optical ray tracing for a system with an axicon combined with a focusing lens. An annular beam is formed at the lens focal plane. The normalized intensity distribution at this position is plotted in false colors at the right. Images credits, Camilo Florian.

On a similar trend, a *Vortex beam* can also produce the same doughnut-shaped intensity. However, its origin extends more than regular intensity distribution changes. In this case, it is the light phase that is customized to change along the beam propagation direction, following a twisted phase trajectory that results in the cancellation of certain light waves, producing a region with zero intensity at the center, as it can be seen in Figure 4 [Spinello, 2016]. Experimentally, it can be produced with a Gaussian beam passing through an optical device that modifies its phase, changing its effective intensity distribution [Wang, 2018]. Different methods can be used for the creation of such beams, requiring in some cases additional optical elements such as phase plates [Guo, 2020], q-plates [Anoop, 2014] [Nivas, 2017], spatial light modulators [Allegre, 2012], deformable mirrors [Gong, 2014], birefringent liquid crystals [Ge, 2017] and s-plates [Tang, 2021], or by using different oscillation modes generated inside a laser cavity.

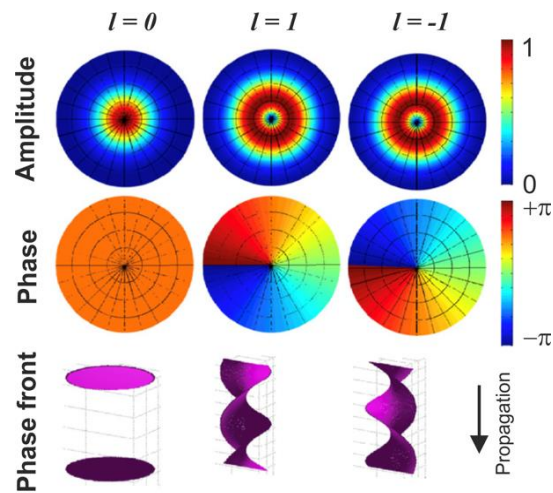

Figure 4. Examples of Laguerre-Gauss modes and superpositions of opposite-sign modes '*l*'. First row: amplitude distributions of the electric field. Second row: phase distributions. Third row: shape of the phase front as it propagates. Reprinted from [Spinello, 2016], Copyright (2016), with permission from Elsevier.

From the point of view of the spatial intensity distributions, Annular beams and these particular Vortex beams with a doughnut shape can be equivalent. The radical difference could be found when the polarization of the Vortex beam is modified, in which case the beam is known as a Vector-Vortex beam. When a material interaction is triggered by certain polarization from the Vortex beam, the interaction from an Annular beam compared with a polarized Vortex beam results to be substantially different, as it is illustrated with two irradiation examples in Figure 5. In this figure, the modification with a regular annular beam produced a ring-shaped modification that is not modified in the center, where the intensity minimum is located. In contrast, the modification with a vector-vortex beam could induce the formation of superficial structures in the irradiated area, whose orientation depends on the local polarization of the beam. In this particular case, the vector-vortex beam has azimuthal polarization, producing concentric micrometric ring-like structures perpendicular to the polarization. Vortex beams have been extensively used in a wide variety of applications [Shen, 2019], including the generation of optical traps [Gao, 2017][Gong, 2018] and the laser processing of materials for biomimetic applications [Skoulas, 2017].

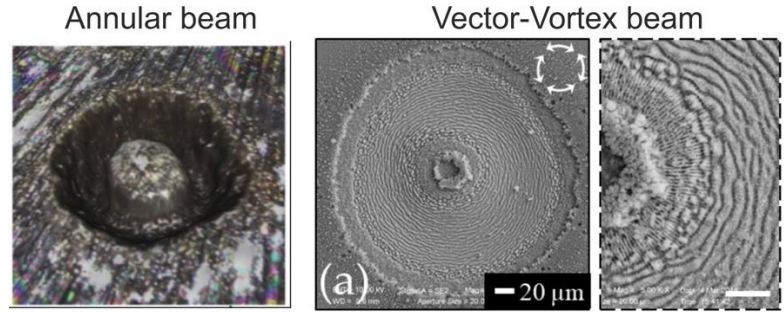

Figure 5. On the left, an optical image from the irradiation of a silicon nitride sample with an Annular beam distribution, produced at a position before the focal plane of an optical system that focalizes an Airy pattern, as illustrated in Figure 2. Reprinted from [Nasrollahi, 2020] Copyright (2020), with permission from Elsevier. On the right, an SEM micrograph shows the modification on a silicon sample using a Vector-Vortex beam with azimuthal polarization as indicated by the white arrows. The fabricated microstructures are defined by the polarization direction. Image adapted with permission from [Nivas, 2021] under Copyright license CC BY 4.0.

## 2.3 Non-diffractive Bessel and Airy beams

Certain complex beams are known to be non-diffractive or quasi non-diffractive, resulting in an ability to reconstruct around obstacles and maintain a focused intensity distribution over long axial distances. However, in reality, non-diffractive beams experience some degree of diffraction due to the finite optics size and beam energy, but their experimental equivalent still offers their characteristic advantages in terms of extended confocal parameters.

Among the non-diffractive beams, the **Bessel beam distribution** exhibits a high-intensity central peak surrounded by less intense concentric rings. Depending on the way these beams are generated, the axial region where the intensity distribution remains approximately constant could range from microns up to several millimeters depending on the optical configuration. Modifications achieved by this type of beam are found primarily in laser cutting and laser marking applications, where it is possible to process millimeter-thick substrates with only one single laser pulse [Duocastella, 2013a].

For this particular beam distribution, the electric field follows the zeroth-order Bessel function of the first kind $J_0$. This description corresponds to a high-intensity central peak that is the product of a superposition of plane waves with wave-vectors forming a cone around the propagation axis that can be exploited for materials processing over large axial distances [Courvoisier, 2013]. Experimentally, the equation that describes such beams where $r^2 = x^2 + y^2$ represents the radial distance from the center and $k^2 = k_z^2 + k_r^2$ is the wave vector, is given by:

$$E(r,t) = A_0 \exp(ik_z z) J_0(k_r r). \qquad \text{Eq. 2.3-1}$$

An approximation of a Bessel beam distribution can be produced experimentally using a Gaussian beam in combination with an aperture [Durnin, 1987], an axicon [Nguyen, 2020][Stsepuro, 2020], a digitally controlled refractive device [Chattrapiban, 2003], and tunable acoustic gradient (TAG) lenses [McLeod, 2006] or other methods [Duocastella, 2012]. Perhaps, the most cost-efficient way to generate a Bessel beam distribution is by using an axicon. However, it requires delicate alignment in order to avoid distortions. Table 1 displays the plots of a Bessel distribution using a wedge axicon angle $\alpha = 2°$ that corresponds at the same time to the angle between the *z* and *r* components of the wave vector *k* [Fontaine, 2019].

Similarly, useful parameters for material processing applications in the case of Bessel beam distributions are the Rayleigh length $z_R$ and the diameter of the central lobe $d_f$ [McGloin, 2005]:

$$z_R \approx \frac{D}{2(n-1)\tan(\alpha)}, \qquad \text{Eq. 2.3-2}$$

$$d_f \approx \frac{2.405\alpha}{k(n-1)\tan(\alpha)}. \qquad \text{Eq. 2.3-3}$$

*n* is the refractive index of the axicon material, and *D* is the diameter of the beam entering the axicon. It is important to note here that in the case of Bessel beams generated with an axicon, both the Rayleigh length and the minimum size of the central lobe depend strongly on the axicon wedge angle. Also, when compared with the values that are possible to reach with a Gaussian beam, the Rayleigh length, in this case, is several orders of magnitude longer, while keeping a rather constant and small central lobe diameter. This characteristic is exploited in materials processing applications that require elongated energy volumes in the axial direction for the control of particles in optical trapping systems [McLeod, 2008][Gong, 2018] or the ablation of thick substrates with a single laser pulse [Courvoisier, 2013][He, 2017]. Figure 6 contains an SEM image from a transversal cut on an irradiated Silicon sample where a conventional Bessel beam was used. In this case, it is possible to see the modification from the main Bessel beam lobe, as well as the three first surrounding annular rings that propagate vertically downwards. For comparison with a regular Gaussian beam, the intensity profile and propagation of a Bessel beam are included in Table 1.

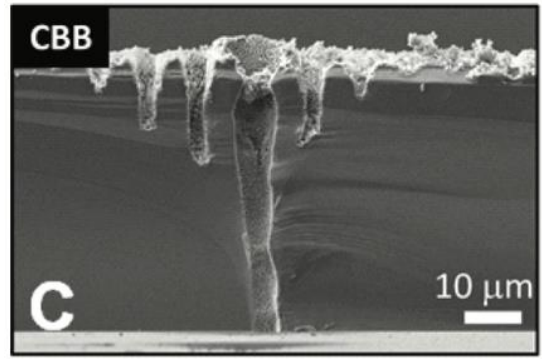

Figure 6. Material modification with a conventional Bessel beam 'CBB' on a Silicon substrate. Image adapted with permission from [He, 2017] under Copyright license CC BY.

Another example of a non-diffractive beam is the ***Airy beam distributions***. Here, the beam appears to be following a curved path along the propagation axis. In this case, the intensity maxima transversely 'accelerate' during their propagation, resulting in a curved trajectory. The intensity distribution of an Airy beam consists of a main lobe that holds the intensity peak preceded by a series of adjacent and gradually less intense peaks that propagate without diffraction. Normally, if there are no changes in the refractive index of the propagating medium, the intensity centroid of an optical beam should move in a straight line [Efremidis, 2019]. This particular property of Airy beams allows the beam to propagate around physical obstacles and reconstruct the intensity distribution at later axial positions. For this reason, Airy beams are often called self-healing beams. This property can be exploited for light propagation in turbulent media without compromising the final beam intensity distribution.

An example of an Airy beam generated by using a Gaussian beam ($w_0$ = 3 mm) passing through a cubic phase mask is displayed in Table 1, which includes the intensity distributions at the XY plane and its propagation along the Z axis. It is important to note that the intensity maxima exhibit a curved trajectory and its length is significatively larger than its Gaussian or Bessel counterparts. In this case, the beam propagation is plotted over a distance of 10 cm. In the particular case of materials processing, it is important to note that significant energy losses are present, especially when compared with the incoming Gaussian laser beam, which makes the use of this type of beam for limited materials processing applications, including the modification of glass [Gecevičius, 2014] and diamond [Mathis, 2012] to name a few. A transverse view of a sample irradiated with an Airy beam after 1, 2 and 5 passes over a silicon sample is shown in Figure 7. The resulting material modification follows the distribution of the beam intensity maxima, which curves along the propagation axis, consequently modifying the material following a curve.

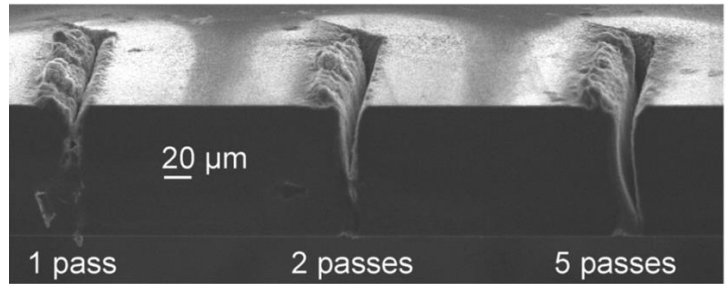

Figure 7. Evolution with the number of passes of the produced machining of curved trenches in silicon using accelerating Airy beams. Reprinted from [Mathis, 2012], with the permission of AIP Publishing.

Other applications for this type of beam can be found in telecommunications [Zhu, 2018], optical trapping systems [Zheng, 2011] and imaging techniques [Guo, 2021], to name a few. A complete set of applications on materials micromachining of materials with Bessel and Airy beams could be found in Chapter 15 "Non-standard light for nanostructuring".

## 2.4 Holography approaches

All the previously mentioned intensity distributions can be achieved by holographic approaches. Here, an initial beam, usually with a Gaussian distribution, interacts with a mask or, more commonly, a digital optical device such as a spatial light modulator or a digital micromirror array [Forbes, 2019] for a user-defined change of its amplitude and/or phase. The general idea is the generation of additional wavefronts that will ultimately interfere at a certain plane along the propagation axis, generating an interference pattern that will contain the desired beam distribution. The implementation of computer-controlled digital devices allows for flexibility in the creation of intensity distributions, enabling the creation of complex forms, as can be seen in Figure 8, where spatially structured beams suitable for marking the surface of a material were generated using a spatial light modulator [Zupancic, 2016][Liu, 2018]. In short, this flexibility allows holographic approaches to create virtually any beam distribution. However, there are some limitations in terms of the maximum energy available after the beam shaping process as well as considerations for the damage threshold for the holographic generating device. Despite this, the use of holographic approaches for materials processing is sometimes desired, because with these methods it is possible to create multiple laser beams that can be used for parallelized processing, increasing the overall processing speed [Cugat, 2013][Martinez, 2015].

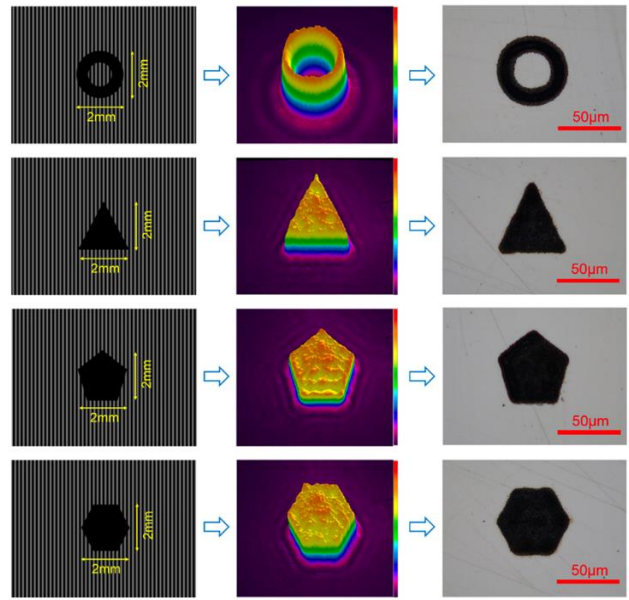

Figure 8. Spatially structured beams formed from different mask spatial distributions by the incidence of a Gaussian beam on a spatial light modulator (left) with the corresponding image of the beam on a CCD camera (center) and the corresponding modification on the surface of a stainless-steel polished sample. Reprinted from [Liu, 2018], Copyright (2018), with permission from Elsevier.

| Description | Intensity distributions | Propagation |
|---|---|---|
| **Gaussian**: Intensity peak located at the center from where the intensity decreases radially. | (A), (B) | (C) |
| | The minimum modified feature is achieved laterally and axially only when the beam's high-intensity location matches the position of interest. Oftentimes, positioning is challenging, especially with high NA optics. | |
| **Annular-Vortex**: High-intensity region that follows a donut shape with an intensity minimum at the geometric center. It could be produced by intensity distribution changes of light phase modifications. | (A), (B) | (C) |
| | The modification in the material follows a ring shape. Contrary to Gaussian beams, the material will not exhibit modifications at the central region of the irradiated area. The beam propagation in this case depends on the way it was generated. In this particular case, the beam generation followed the optical system showed in Figure 3. | |

| | |
|---|---|
| **Bessel**: High-intensity central lobe surrounded by concentric rings with less intensity. Propagation could be several micrometers while keeping a rather constant intensity distribution. | 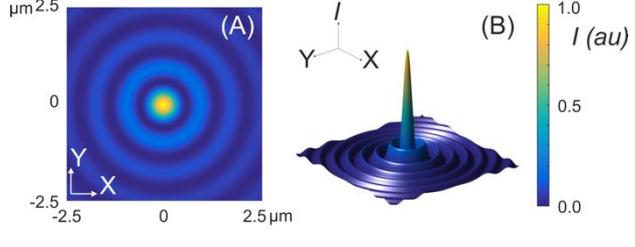 |
| | The modification occurs over longer axial ranges compared to conventional processing with Gaussian beams. Depending on the energy used and the material bandgap, it would be possible to produce modifications only with the central lobe. |
| **Airy**: Intensity peak follows a curved trajectory along the propagation axis. Similarly, as with Bessel beams, propagation could range several millimeters. | 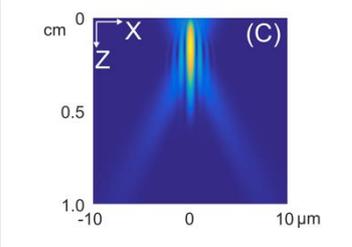 |
| | Contrary to Bessel beams, materials modified with this distribution will show curved modifications along the propagation direction. It provides longer axial ranges for processing compared to regular Gaussian beams. |

Table 1: Intensity distributions and propagation along the Z-axis of the different beam distributions of *Section 2*. The plots for each beam intensity distribution are simulations of (A) Top view of the XY plane, (B) 3D representation, (C) XZ projection showing the propagation along the Z axis. Images credits, Camilo Florian.

# 3 Near-field approaches

Near-field optics describes and exploits light-matter interactions at a nanometer scale, where the optical field is evanescent rather than propagating. Thus, it is not subject to the classical diffraction limit, allowing outstanding spatial resolution and enhanced intensity that can be used in applications for plasmonics, microscopy, nano-photonics and nano-fabrication. In general, any monochromatic light can be represented by an electric field *E(r)* with spatially dependent amplitude *A(r)* and phase *ϕ(r)* of the vector components of the field, following Eq. 3-1:

$$E(r) = A(r)e^{i\phi(r)} = \int_{-\infty}^{\infty} E_k e^{ik.r} \, dk, \qquad \text{Eq. 3-1}$$

where *r* represents the spatial position, and $E_k$ is the complex amplitude of the wavevector *k*:

$$k = \frac{2\pi n}{\lambda} = \sqrt{k_x^2 + k_y^2 + k_z^2}. \qquad \text{Eq. 3-2}$$

The real part of the refractive index is represented by *n* and the light wavelength by *λ*. The Z component of the wavevector has a non-zero imaginary component when $k_x^2 + k_y^2 > k_0$, and therefore, any imaginary components in Eq. 3-1 will have an exponentially decaying amplitude with an increase in distance. The near-field that is localized within less than a wavelength distance from the surface of interest evolves towards propagating far-field radiation that contains information from the material at the surface, with resolutions below the diffraction-limited maximum resolution of the far-field [Bazylewski, 2017].

In this subsection, we briefly review two well-established techniques to manipulate the near-field in the nanoscale, including plasmonic approaches such as scanning near-field optical microscopy (SNOM) probes, and dielectric-based approaches such as the use of optically trapped dielectric microspheres.

## 3.1 Plasmonic based approach: SNOM probes

In scanning near-field optical microscopy (SNOM), the generation of near-field light can take place by using two different sets of methods. The first general set is aperture-based approaches that consist in the use of a probe with an aperture whose size is smaller than the laser wavelength used to illuminate the sample. Here the near-field radiation is scattered back to be detected as far-field radiation following a configuration as the one shown in Figure 9(A). A second general set of methods consists in aperture-less approaches, where the near-field light is emitted from the surface due to external illumination that is scattered off of a sharp tip to be detected likewise in the far-field [Bazylewski, 2017]. The use of the same principle as the one for imaging structures with nanometer resolution, but for the modification of materials, has been demonstrated with different photosensitive materials, polymers, metals and semiconductors, to name a few [Roszkiewicz, 2019]. Here, when the near-field enhancement overcomes the required ablation fluence for a given material, it is possible to produce permanent modifications with nanometer resolution and high reproducibility. Figure 9(A) shows a system that incorporates an ***SNOM aperture-based probe*** to generate the near-field radiation over a photosensitive polymer material. The modifications produced are included in Figure 9(B), where a set of craters were characterized via AFM. From the topography data, profiles that reveal the depth of the modifications are extracted along the line of craters, showing modifications of ~500 nm in diameter and ~75 nm in depth.

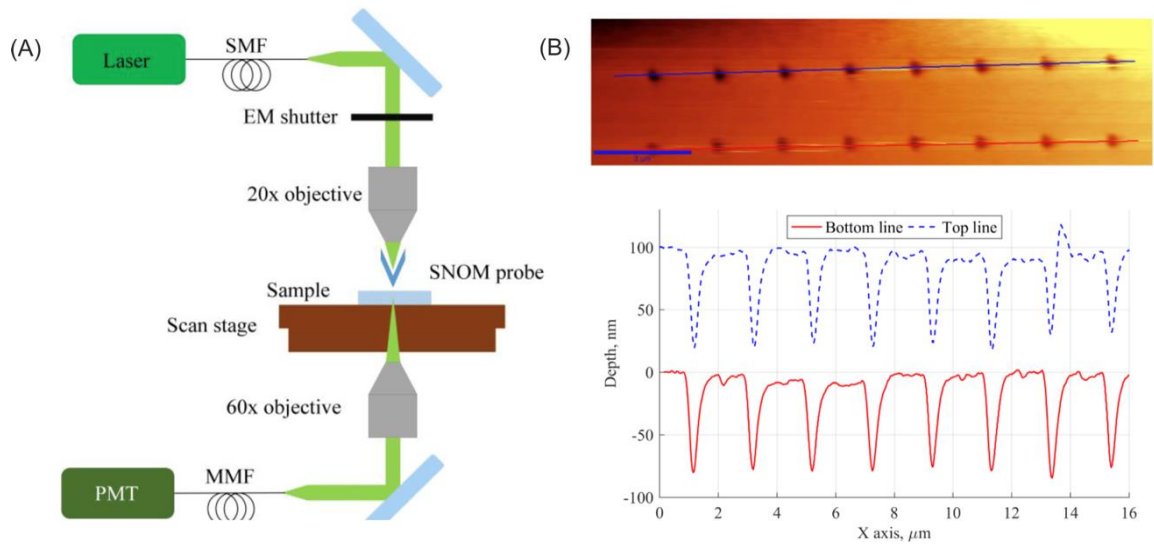

Figure 9. (A) Schematic representation of the SNOM experimental setup. (B) AFM topography data of the irradiation experiments performed on a photoresist deposited on quartz, including a cross-sectional profile along the marked blue and red lines. Image adapted with permission from [Roszkiewicz, 2019] under Copyright license CC by 4.0.

Additional methods to shape light in the near-field can profit from plasmonic interaction based on the use of different geometries of antennas, in which the near-field enhancement is larger compared to values reached with circular geometry probes. Among these configurations, there are several approaches based on *plasmonic lensing structures* initially designed for imaging applications, that consist in concentric ring gratings with defined refractive indexes for the detection of the near-field at the vicinity of the structures of interest offering nanometer resolutions [Chaturvedi, 2010] [Fang, 2012]. For materials processing, due to the extremely small cavity sizes of plasmonic lenses, the intensity enhancement could be up to 100 times higher than the laser radiation impinging on the structures. When the appropriate system is used, the energy could be focused in the near-field to spots smaller than 100 nm [Sriturawanich, 2008].

Alternatively, **Bow-tie antennas** are widely employed in plasmonic structures for materials processing. When combined with femtosecond laser pulses, they generate highly confined intensity peak distributions suitable for nanopatterning applications. The advantage is attributed to a field concentration in the gap between the ridges of the antenna. Figure 10 includes SEM micrographs that display the bow-tie antenna fabricated on the tip of an AFM cantilever. This particular system was used in the femtosecond-based nanopatterning of a positive photoresist Shipley (S1805) with features as small as 24 nm in width, following at the same time a curve trajectory (Figure 10(B)). The enhancement is finally produced by the propagation of the fundamental electric-magnetic field TEM10 mode. Ultimately, the intensity variation is different depending on the geometry of the SNOM probe [Murphy-dubay, 2008][Guo, 2010][Kinzel, 2010][Kumar, 2011].

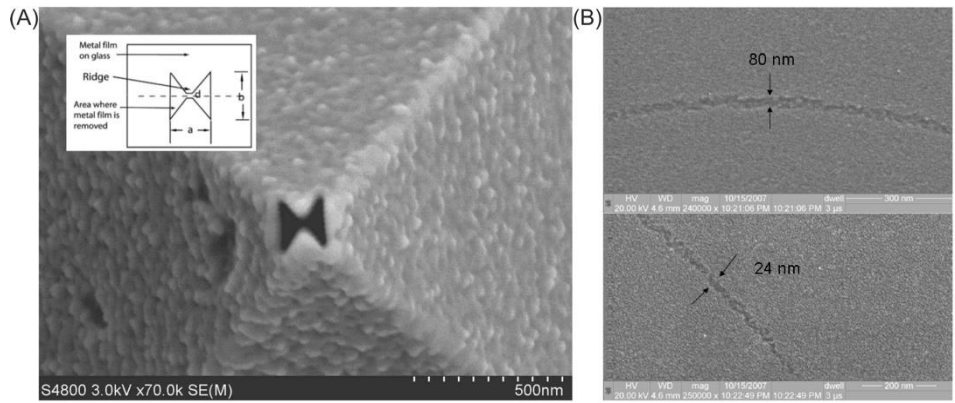

Figure 10. (A) SEM image of a bow-tie aperture on an AFM probe with an inset of idealized geometry. (B) SEM of circle pattern showing top and side sections of a circle. Reproduced from the Open Access paper [Murphy-dubay, 2008] ©2008 Optical Society of America.

## 3.2 Dielectric based approach: Optical-trap assisted nanopatterning

One of the first experimental demonstrations on the use of near-field enhancements produced by certain particles was reported during the development of techniques for cleaning semiconductor surfaces with laser light. The idea was to use loosely focused laser beams to generate a catapulting effect to detach particles from the surface of expensive substrates [Imen, 1991]. Unfortunately, for its original application, this catapulting effect led to unwanted substrate permanent nanometric modifications. Later, this idea of exploiting the field enhancements was developed into particle- and tip-assisted nanofabrication techniques in which different and numerous research groups have contributed to the understanding of the interaction of matter with light at these extremely small scales [Leiderer, 2000][Luk'Yanchuk, 2003][Kühler, 2009][Wang, 2010][Surdo, 2021].

As a particular example of these techniques, the use of *dielectric microspheres* positioned conveniently near a substrate could exhibit the near-field radiation enhancements for processing materials in the nanoscale. Here, radiation coming from a laser source passing through the microsphere can be focused onto very small focal volumes, enhancing the near-field between the microsphere and the substrate of interest. If the intensity reached is high enough, the material can be permanently modified, exhibiting subwavelength spatial resolution. In this particular case, the dielectric microlens behaves as an aperture that allows the subsequent focusing of light, as it is shown in Figure 11. In this figure, an optically trapped dielectric microsphere in water is used to enhance the near-field radiation in the vicinity of a polyimide film and a silicon substrate. The lateral positioning of the microsphere relative to the irradiated sample is performed by piezo-actuated stages. The axial distance of the microsphere is kept constant as a product of the double-layer repulsive potential between the microsphere and the substrate, as well as an optical force pushing the microsphere towards the substrate. The constant balance between attraction and repulsion preserves the same microsphere-substrate distance at all times without feedback control mechanisms. This is possible due to the nature of the Bessel beam optical trap that allows for a large axial trapping region where the optical force on the microlens is practically constant. Depending on the irradiation conditions, it is possible to produce craters in the material by ablating the material in selected positions (Figure 11(D)), or to induce material swelling for the formation of nano-cones (Figure 11(E)).

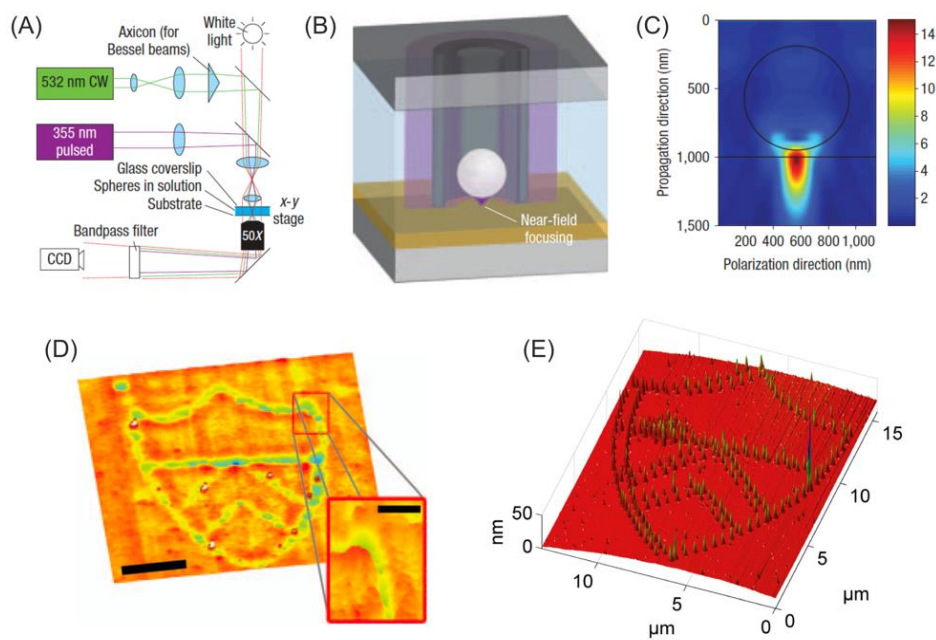

Figure 11. (A) Schematic of the experimental setup including a trapping laser (green) and a processing laser (purple). The white light is used for real-time video acquisition with the 50× objective and the CCD camera. (B) Sketch of the irradiation process with an optically trapped sphere near the polyimide surface (orange) in the trapping laser (green). The processing beam (purple) covers an area larger than the microsphere, but it only reaches fluences above the threshold for modification in positions directly beneath the microsphere. (C) FDTD simulation of the intensity passing through a 0.76 µm microsphere in water above the polyimide substrate. The plane shown is parallel to the propagation and electric field polarization directions. (D) AFM image of a Princeton University logo nanopatterned on the polyimide layer. In this case, a fluence

of 3mJ/cm$^2$ was used to produce trenches by overlapping consecutive laser pulses. Main scale bar, 2 mm; inset scale bar, 1 mm. (A-D) Reprinted with permission from [McLeod, 2008], Copyright 2008, Springer Nature. (E) AFM image of a Princeton University logo made of nano-cones on a silicon sample produced at a fluence of 7.3 mJ/cm$^2$ with a polystyrene microsphere. Reprinted with permission from [Chen, 2017], Copyright 2017, Laser Institute of America.

A summary of the intensity distributions used in the above near-field approaches is arranged in Table 2. From here, different approaches based on the use of dielectric objects have emerged for applications in materials processing and imaging [Chen, 2020]. A complete description of available techniques and historic development of applications that use near-field radiation for the modification of materials is included in Chapter 11 on "Optical near-field structuring".

| Description | Intensity distributions |
|---|---|
| **SNOM probes**: Here, the field enhancement is produced between the material and the probe for distances smaller than the laser wavelength used. The intensity forms a narrow filament in the near-field that disperses in the far-field. | 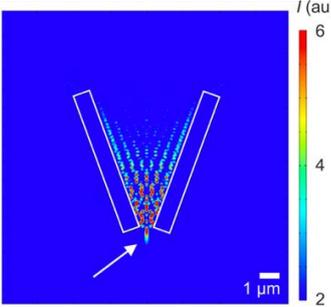 |
| | The field enhancement exploited for materials processing is produced by approaching the SNOM probe close to the surface of the material of interest. In that way, the resolution enhancement is combined with an intensity increase that allows the machining of materials with high spatial resolution. The dimensions are considerably smaller than the Gaussian counterparts. |
| **Bow-tie antennas**: The field enhancement is produced in the tight separation between the triangular structures as a product of plasmon resonances at the structure's surface. | 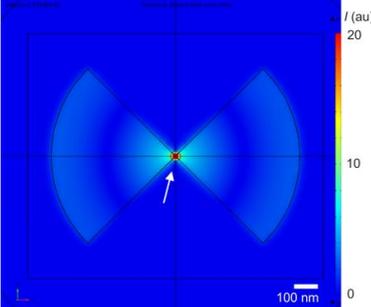 |
| | The field enhancement in bow-tie antennas configurations can be used for nanopatterning applications over extended areas when the antenna is mounted in a scanning-compatible device, as shown in Figure 10. |

| **Dielectric microsphere**: the intensity distribution used for materials processing is formed after the dielectric microsphere. The voxel is known as a photonic nano-jet, due to its small dimensions. | 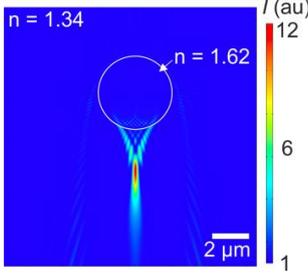 |
|---|---|
| | The final intensity distribution dimensions used for materials processing depend on the wavelength used, the refractive index ratio between light propagation media/microsphere, and the distance between the microlens surface and the sample of interest. |

Table 2. Intensity distributions and propagation along the Z-axis of the three different beam distributions from *Section 3*. Images credits, Camilo Florian.

# 4 Time-dependent structured profiles

In addition to the static structured light approaches discussed in the previous section, there exist ultrafast beam structuring methods that exploit system or material properties that evolve over time. Depending on the specific application, it is possible to induce (i) transient absorption regimes caused by the action of two adequately separated beams in time or space, (ii) interference between two or more laser beams, or (iii) rapid variation of focus produced by an ultrafast varifocal lens, all of which are aimed towards the production of high spatial resolution features. In the following subsections, we will discuss how these time-dependent structured profiles can be used in different laser-based processing techniques.

## 4.1 Dual-laser beam processing 'DLBP' and Dual-laser additive manufacturing 'DLAM'

Multiple beams enable us to create material modifications that are not possible to produce with regular single beam processing. The resulting interaction between more than one beam extends the benefits for different materials processing applications towards higher spatial resolutions either from a subtractive or additive process.

***Dual-laser beam processing (DLBP)*** of materials follows a subtractive approach based on the same principle as the super-resolution fluorescence microscopy technique STED [Hell, 1994]. In DLBP, two different laser beams irradiate the sample at the same position, one with Gaussian spatial and temporal distribution and pulse energy above the material's bandgap to promote electrons from the valence band to the conduction band, and a second one with donut-shaped spatial profile and Gaussian temporal profile and pulse energy lower than the material's bandgap to promote the depletion of the previously excited electrons back to the valence band, as it is illustrated in Figure 12. Due to the donut-shaped spatial distribution, after the depletion process takes place, in the central part, unirradiated by this second beam, excited electrons are located in a significatively smaller region. If a third Gaussian pulse reaches the sample when this transient regime is present, the excited electrons in the conduction band can be easily ablated with a wavelength that would otherwise be transmitted by the sample, reducing the final feature size. This induced transient state occurs within the duration of the laser pulses and can be controlled by shifting the pulse delays in a controlled way [Wenisch, 2021]. It was determined that the best inter-pulse delay for the realization of the smallest features corresponds to +5 ns (Figure 12(C)), since a combination of heat and excited carriers increases absorption for the implemented laser wavelengths.

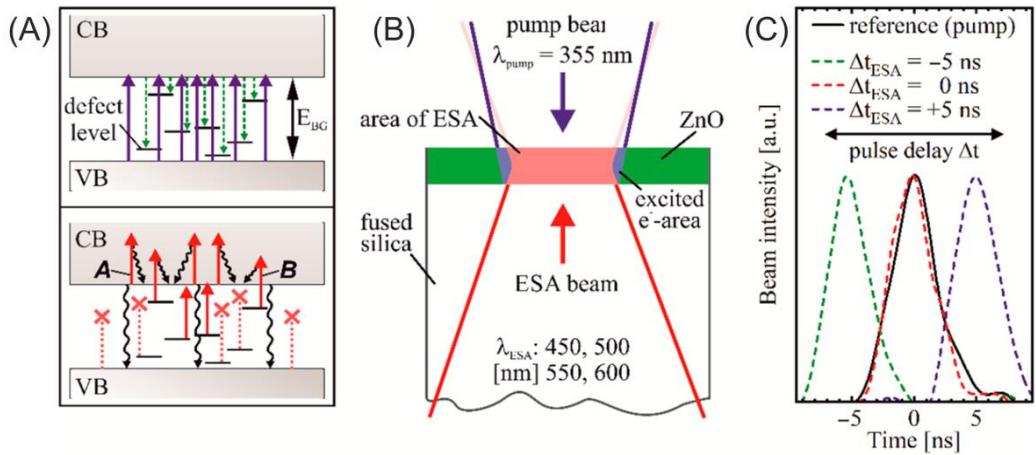

Figure 12. (A) Illustration of the excited state absorption (ESA)-based ablation process. The upper part shows the excitation from the valence band (VB) to the conduction band (CB) and the bottom part shows the ESA in solid arrows. (B) Sketch of a 1mm thick sample cross-section (white) covered by a thin layer of ZnO (green). The pump beam impinges the sample from the top and the ESA beam from the bottom, as indicated by the arrows. (C) Pulse delays (dashed lines) and their temporal overlap with respect to the pump pulse (solid black line). Image adapted with permission from [Wenisch, 2021] under Copyright license CC by 4.0.

Additive manufacturing techniques can also take advantage of multiple laser beams to modify and control the material processing. For instance, in laser powder bed fusion, the fabrication of metallic 3D objects is achieved by selectively melting irradiated areas in a bottom-up configuration following a layer-by-layer approach. In order to increase the fabrication speeds in this type of additive manufacturing technique, parallelization using ***dual-laser additive manufacturing***

***(DLAM)*** approaches is often preferred [Yap, 2015]. However, different variables should be considered for a correct application of the approach, including the geometrical distribution of the laser beams with respect to the final fabricated feature, and the thermal processes associated with the interaction of different laser beams as heat sources. By controlling these parameters, it is possible to exploit regions of already melted material to modify how the melted pool will cool down and solidify. Following this approach, it is possible to modify and control the thermal evolution of the fusion process and fabricate morphologies and structures that otherwise would be difficult to produce with conventional single beam powder bed melting approaches [Zhang, 2020]. Figure 13(A) shows a schematic layout of the geometry used to create two parallel metallic lines. When the hatch distance between them is reduced, the two lines coalesce into an intermediate reproducible state of periodic structures of melted material, as it can be observed in Figure 13(B).

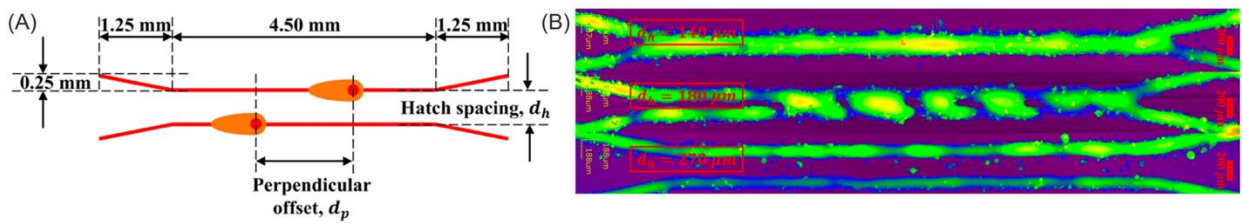

Figure 13. (A) Schematic diagram defining hatch spacing and perpendicular offset in the experiment. The red lines will correspond to lines that protrudes from the surface as a product of melting a metallic powder bed. (B) Lines with different hatch distances are produced and characterized via laser scanning confocal microscopy. Reprinted from [Zhang, 2020] Copyright (2020), with permission from Elsevier.

Interestingly, depending on the perpendicular offset, for two identical beams, it is possible to observe two types of material coalescence; for no perpendicular offset, the coalescence follows a heat-to-head fashion, where each individual molten pool interacts with each other forming material bulges with a certain periodicity but no preferential orientation. In the second situation, when the perpendicular offset is increased, the interaction follows a head-to-tail fashion, resulting in a clear orientation of the periodic structures of the molten pools [Zhang, 2020].

## 4.2 Direct laser-interference patterning 'DLIP' and Laser-induced forward transfer 'LIFT'

When the interaction between more than one beam occurs within a determined spatial region during a short timeframe, it is possible to induce interference between the implicated beams.

Similarly, as with dual-laser beam processing approaches, this interference can be exploited for the processing of materials following subtractive and additive modes.

***Direct Laser-Interference Patterning (DLIP)*** foundation principles are based in a technique called Laser Interference Lithography (LIL) [Seo, 2014], originally used as a maskless large area photolithography technique for the formation of periodic nanometric-scale patterns. DLIP uses higher intensity lasers compared to LIL, and it is also a maskless processing technique that exploits the light coherence to induce spatially-controlled intensity patterns for the fabrication of micro- and nanometer-scale periodic structures. The process consists in the interference of two or more coherent laser beams on the surface of the material of interest. In such cases, the intensity pattern periodicity depends on the wavelength used and the incident angles of the incoming laser beams. The shape of the resulting pattern can be defined by changing the relative phase between the incident beams, and its intensity can be modulated by the polarization of each participant beam. The intensity pattern $I(x,y)$ at the surface for two equal beams with top hat intensity is given by Eq. 4.2-1 [Daniel, 2003], [Peter, 2020]:

$$I(x,y) = 2I_0 \left[1 + \cos\left(\frac{4\pi \sin\theta}{\lambda}x\right)\right], \qquad \text{Eq. 4.2-1}$$

where $I_0$ is the peak intensity of each of the two incident laser beams, $\lambda$ is the laser wavelength, and $\theta$ is the beam's angle of incidence with respect to the surface. For the case of a Gaussian spatial distribution, the intensity pattern is the following:

$$I(x,y) = 4I_0 \exp\left[\left[-2\left(\frac{4x^2}{d_x^2} + \frac{4y^2}{d_y^2}\right)\right] \cdot \left[1 + \cos\left(\frac{4\pi \sin\theta}{\lambda}x\right)\right]\right]. \qquad \text{Eq. 4.2-2}$$

For the interference system, it is also possible to calculate the expected period $\Lambda$ with Eq. 4.2-3:

$$\Lambda = \frac{\lambda}{2\sin\left(\frac{\theta}{2}\right)}. \qquad \text{Eq. 4.2-3}$$

An example of the calculated and measured intensity patterns is included in Figure 14, where it is also possible to see the ablation on stainless steel using two different pulse energies.

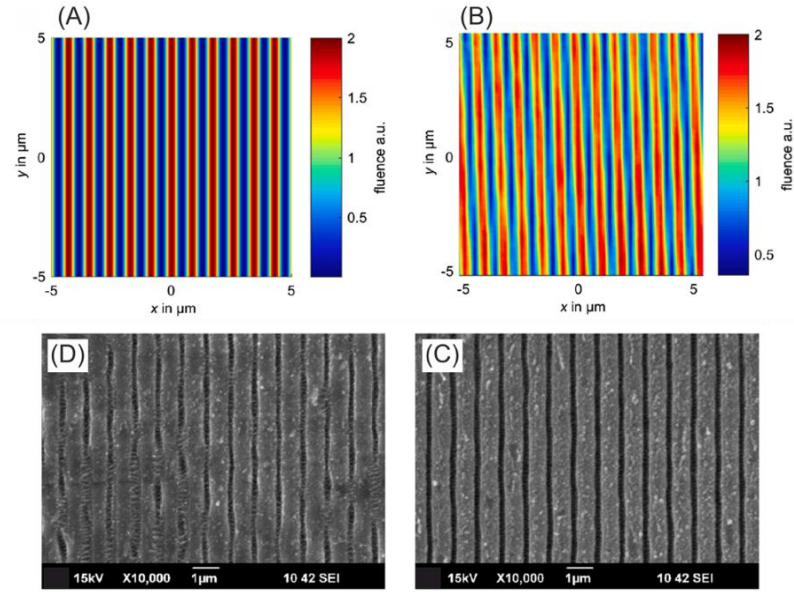

Figure 14. Calculated (A) and measured (B) intensity patterns at the sample surface. The corresponding irradiations in stainless steel are presented as SEM micrographs in (C) for pulse energy $E_p = 7.2\,\mu J$ and (D) for $E_p = 9\,\mu J$. Image adapted with permission from [Peter, 2020] under Copyright license CC by 4.0.

Increasing the number of beams participating in the interference produces more complex intensity patterns; for instance, in planar configurations, three beams can produce a two-dimensional dot pattern, while four beams can be used to produce a line pattern with thinner and higher peaks and thicker areas of low intensity than a two-beam pattern [Lasagni, 2005],[Rebollar, 2020]. An example for non-planar beams is included in Figure 15(A), producing an intensity pattern that corresponds to multiple spots separated longitudinally by a period $\Lambda$. The product of the irradiation on titanium with this intensity profile can be seen in the SEM micrograph of Figure 15(B), where multiple spots were irradiated to cover a whole area. When the irradiation takes place with ultrashort pulsed lasers, the formation of low and high spatial frequency LIPSS (LSFL and HSFL) can also be induced, as shown in Figure 15(C). Detail applications for the use of these hierarchical structures will be extensively discussed in Chapter 23 "Ultrarapid industrial large area processing of surface nanostructures".

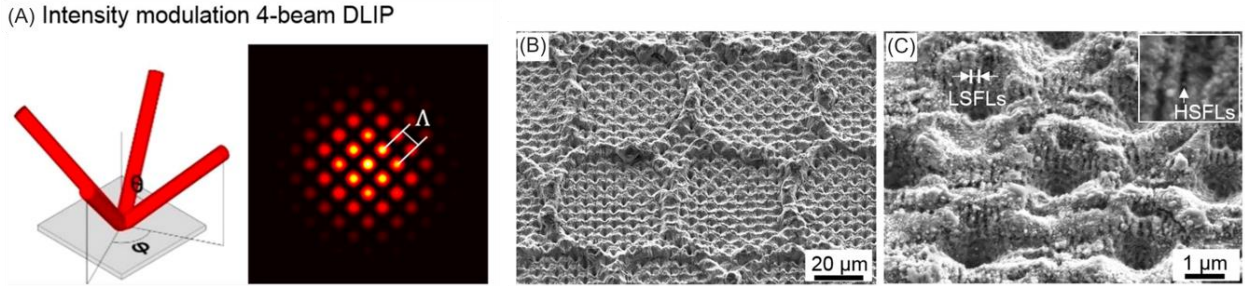

Figure 15. (A) Illustration of beams impinging on the sample simultaneously from different angles and the resulting calculated intensity pattern from black (no intensity) to yellow (high intensity). (B) Hierarchical surface structure on titanium produced with the intensity pattern shown in (A). (C) High magnification image showing also the LIPSS features (LSFLs in the image and HFSLs in the inset). Image adapted with permission from [Zwahr, 2019] under Copyright license CC by 4.0.

Printing of materials under a regular *laser-induced forward transfer (LIFT)* process occurs when a focused laser beam is irradiated over a material of interest that is often disposed as thin layers (solid, liquid or intermediate viscosity pastes) [Arnold, 2007][Serra, 2019]. The irradiation leads to a microexplosion that impulses tiny amounts of the material towards a receptor substrate placed conveniently close by. The resulting feature forms one printed pixel per laser pulse with micrometer resolutions [Florian, 2016][Turkoz, 2018][Turkoz, 2019]. Recently, Direct laser-interference patterning (DLIP) was implemented to generate an intensity pattern distribution corresponding to a matrix of pulses like the one shown in Figure 16. Each individual intensity dot can propel material towards an acceptor substrate while keeping the same spatial distribution even considering the drag resistance under ambient air conditions, as reported in [Nakata, 2021]. In this particular case, the propelled material was disposed in thin films of noble metals, including platinum and gold. The resulting printed pixels in the case of the gold donor substrate correspond to pixels in the shape of sub-micron sized dots that follow the same spatial distribution of the used intensity distribution profile, as shown in Figure 16.

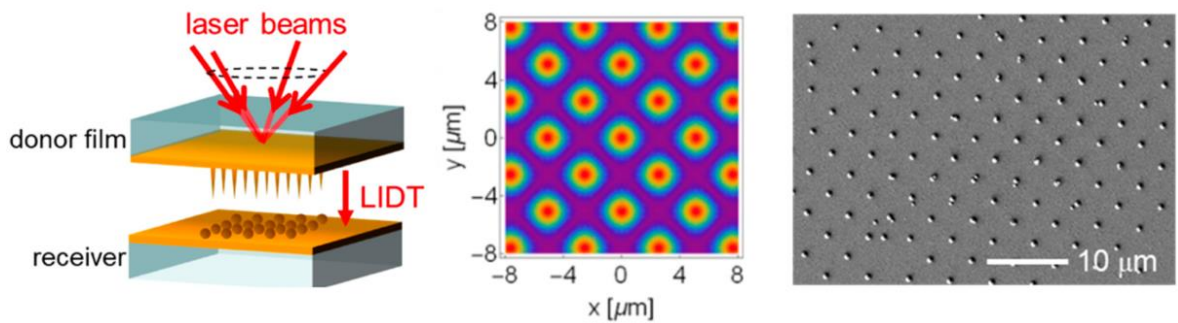

Figure 16. Illustration of the 4-beam interference at the interface of the donor substrate and the thin film of the material of interest. The intensity profile corresponds to the simulated intensity at the same position. On the left, a SEM micrograph displays gold pixels symmetrically spaced printed over a receptor substrate. Image adapted with permission from [Nakata, 2021] under Copyright license CC by 4.0.

## 4.3 Resonant focal scanning

Another important method of beam control for optimal laser processing is the use of ultra-high-speed scanning of the focus in order to create a quasi-line focus or other complex beam shape. There are few technologies able to scan at sufficiently high speed [Kang, 2020]. One type of ultrafast variable focus optic is a resonant scanning device known as a tunable acoustic gradient of refractive index (TAG) lens. These lenses are optical devices that use acoustic waves inside a liquid to modulate the local density, producing spatially and temporally varying refraction index changes that can be exploited for lensing effects [McLeod, 2006][Kang, 2020] (Figure 17A). The oscillation frequency is driven by an external voltage source whose amplitude influences the focusing-defocusing power of the lens. Due to the local refractive index changes, the liquid changes the light propagation path as shown in Figure 17B. The frequency of operation of the TAG lens ranges several hundreds of kHz, and the specific working frequencies depend on the geometry of the piezoelectric itself.

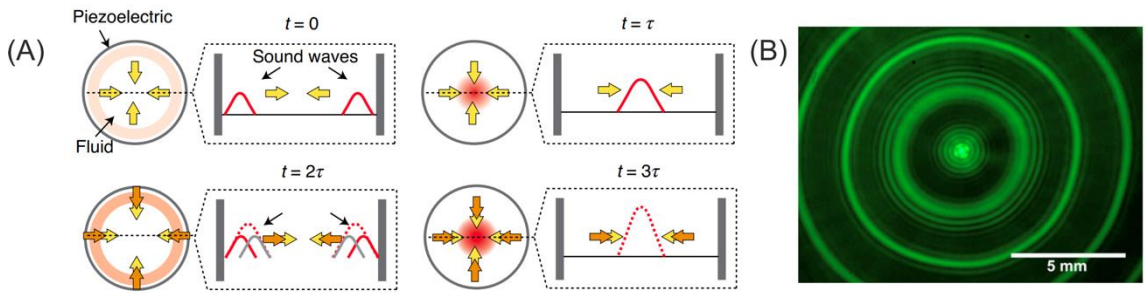

Figure 17. (A) Schematic of the model based on considering the TAG as a resonator. Cylindrical waves travel back and forth in the TAG lens, interfering with each other and producing periodic changes in the lens optical power every 2 τ. Image reproduced with permission from [Duocastella, 2013a]; permission conveyed through Copyright Clearance Center, Inc. (B) Beam distribution measured after 80 cm from the TAG lens using a Gaussian collimated beam as input beam. The piezoelectric frequency is 299.7 kHz. Image reproduced with the permission of AIP Publishing from [McLeod, 2007].

The use of this type of lens presents advantages for the processing of materials, since it is possible to dynamically displace the intensity distribution along the beam axis over longer distances compared with regular direct writing systems. Therefore, it is possible to modify samples without the need of sample repositioning [Duocastella, 2013b][Du, 2021]. Figure 18 shows the intensity distribution for a nanosecond laser marking system that implements a regular marking system and one that incorporates a TAG lens. The range of positions along the beam axis where the intensity is high enough to produce damage in the sample is larger in the case of the system with a working TAG lens. This allows the marking of the surface of a stepped sample without the need for sample repositioning, as shown in Figure 18(C), thanks to the axial scanning performed by the TAG lens.

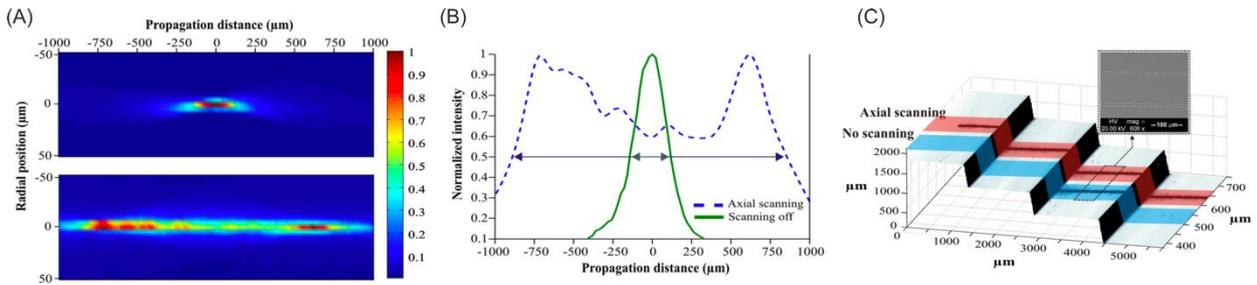

Figure 18. (A) Intensity distribution of the laser beam propagation without and with axial scanning (TAG lens amplitude 10Vpp) after a microscope objective (5×, NA = 0.13). (B) Intensity profiles at the center (radial position = 0 µm) along the axial direction. (C) Machining with and without a TAG lens on a silicon stepped surface. When there is no axial scanning, the system marks the surface only in one step, whereas with the axial scanning provided by a working TAG lens is possible to mark all the surfaces without requiring sample repositioning over an extended axial distance of 2 mm. The inset corresponds to an SEM micrograph with the widths of both modifications at the indicated step. Reprinted from [Duocastella, 2013b], with the permission of AIP Publishing.

Another advantage of using a resonant scanning system is an increase of ablation efficiency provided by the rapidly scanning focal position. Conventional direct-laser writing systems used for the ablation of materials usually rely on the use of higher numerical apertures to produce modification with high spatial resolution. The use of higher numerical apertures means that the axial region where the energy is focused is smaller. When the machining of samples in depth > Rayleigh length is needed, the sample position should be carefully maintained to benefit from the small focal volume; as material is removed, it is important to continue to adjust the location of the laser focus to follow the change in topography of the workpiece [Chen, 2018]. However rapid resonant focal scanning obviates the need for such detailed feedback and control of the sample surface location. Figure 19(A) shows the comparison of the ablation with a nanosecond laser of a square of $200 \times 200$ µm$^2$ on a silicon sample with and without the TAG lens after three consecutive scans. In particular, the vertical profiles show a lack of ablated material in the case of the regular marking system without the TAG lens, whereas in the case with a working TAG lens, it is possible to ablate homogeneously over the whole area. This effect is produced because the scanning lens irradiates the material with different effective numerical apertures (or different focal lengths) per cycle defined by the amplitude of the oscillation. Therefore, the irradiation takes place with different spot sizes that increase material ablation when the fluence is high enough [Chen, 2018]. Optimization of a resonant scanning machining system can be performed by properly selecting the rate between the frequency of the marking laser and the frequency of operation of the scanning element. If the factor between these two parameters is an integer, the focal positions of the irradiated pulses will be the same, and no axial scanning will be induced. For all the other cases, the frequency mismatch will ensure that a pulse reaches the sample at different locations for each pulse interacting with the surface. Figure 19(B) shows an irradiation experiment in a cube of borosilicate glass using the same laser repetition rate (1000 Hz) and different TAG lens operating frequencies [Du, 2021]. Here, the sample is scanned at the same speed laterally and the images are acquired from a lateral perspective. Depending on the machining application, the right number of pulses per cycle could be customized by the proper selection of the TAG lens operational frequency, increasing the number of pulses that effectively reach the sample per TAG lens cycle.

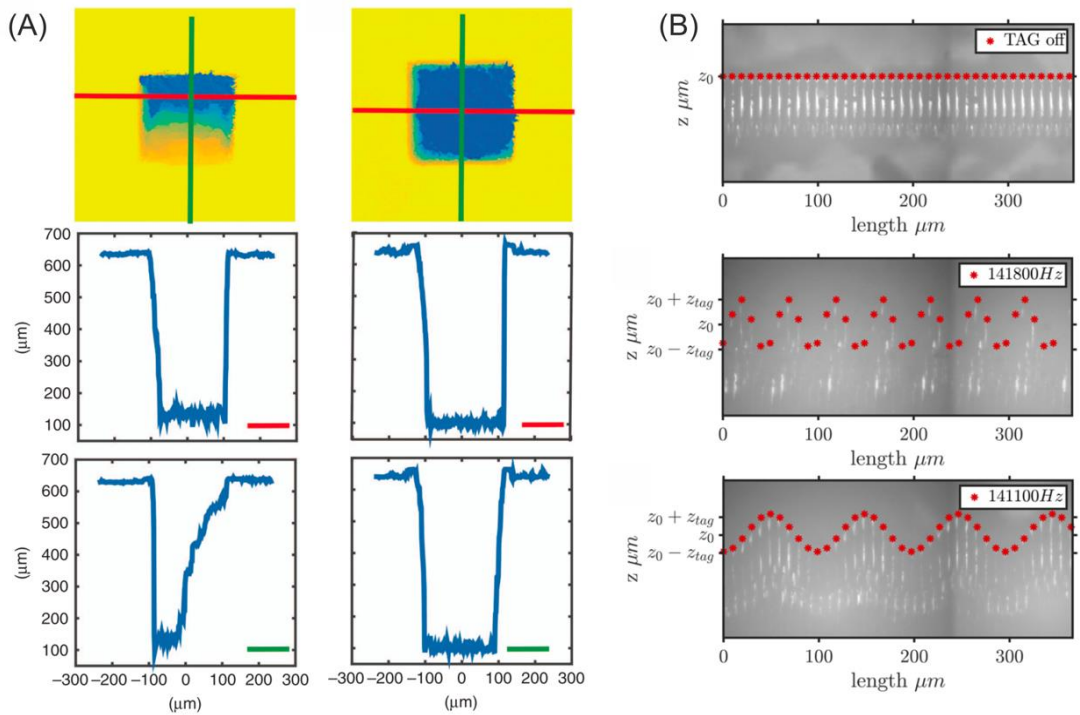

Figure 19. (A) Ablation of 200 × 200 µm² squares on a 500 µm thick silicon sample after three consecutive scans without (left) and with a TAG lens (right). The second row shows horizontal cross-section profiles indicated by the red lines. The third row similarly shows vertical profiles

indicated by the green lines. Image adapted with permission from [Chen, 2018] under Copyright license CC by 4.0. (B) Optical microscopy images from a lateral perspective of femtosecond laser modifications in a borosilicate glass cube without (top) and with an axial scanning frequency of 141,800 Hz and 5 pulses per cycle (middle) and 141,100 Hz and 10 pulses per cycle (bottom). Image reprinted from [Du, 2021], Copyright (2021), with permission from Elsevier.

# 5   Conclusions and perspectives

This chapter reviews techniques for creating and using spatially structured beams from a fundamental perspective, as well as the resulting material interactions that occur. Generally speaking, the three key parameters which influence the final features of laser material processing are the confocal parameter, the beam divergence, and the minimum spot size. These parameters describe how traditional Gaussian beams focus energy into a small voxel and make surface or bulk modifications to materials, following closely the voxel dimensions, however we discuss how it is possible to control and modify these parameters in order to create more complex beam architectures with beneficial performance in laser processing applications. Such spatially and temporally structured beam profiles are classified as varying the spatial intensity distributions in the far-field, the near-field, and temporally varying profiles. We show several examples of how to exploit these time-dependent profiles in subtractive and additive processing techniques. The development and advances of current and future laser-based micro- and nanofabrication techniques rely on our ability to accurately control and modify the optical properties of an incident beam of light and through techniques such as those described in this chapter, great advances in materials processing continue to be made.

**Acknowledgements**

C.F. acknowledges the support from the European Commission through the Marie Curie Individual Fellowship – Global grant No. 844977. Also, to G. Mao for the support and useful conversations.